\documentstyle{article}
\begin{document}
\begin{center}
\begin{Large}
\bf{Kronecker product/Direct product/Tensor product in Quantum Theory}\\
\end{Large}
\bigskip
Z. S. Sazonova\\
Physics Department, Moscow Automobile and Road Construction Institute (Technical University),
64, Leningradskii prospect, Moscow, Russia
\bigskip

Ranjit Singh\\
Wave Research Center at General Physics Institute of Russian Academy of Sciences,
38, Vavilov street, Moscow 117942, Russia Tel./Fax: (+7 095) 135-8234 email: ranjit@dataforce.net \\

\begin{abstract}
The properties and applications of kronecker product\footnote{For simplicity, we will use kronecker product. The other two names: direct product amd tensor product are also similar.} in quantum theory is studied thoroughly.
The use of kronecker product in quantum information theory to get the exact spin Hamiltonian is given. The proof of
non-commutativity of matrices, when kronecker product is used between them is given. It is
shown that the non-commutative matrices after kronecker product are similar or they are similar matrices [9,17,20]. 
\end{abstract}
\end{center}
\section{Introduction}
The use of kronecker product in quantum information theory is used extensively. But
the rules, properties and applications of kronecker product are not discussed in any quantum
theory books [3,5,6,13,14,16,18]. Even books on mathematical aspects of quantum theory are discussing the
properties and applications of kronecker product in very short without any explanations 
of its rules. That is, why we are applying kronecker product left/right to any spin operator,
why not left or why not right only. They are applying kronecker product without any 
explanation. Mathematicians were also not able to give any concreate answer to these 
questions [19] in more generalized way. Nowdays we can find books on algebra [10,20], where the
mathematical aspects are considered in more rigorous and detail. But the books on algebra [1,2,12,15,17]
do not consider the physical aspects of kronecker product, which are very important
in quantum theory. So, we will give answers to all of the above mentioned questions
in this article, which are now very important in quantum information theory to write 
exact spin Hamiltonian. 
\section{Mathematical aspects of kronecker product}
Let $\hat{A}$ and $\hat{B}$ are two linear operators defined in the finite dimensional $L$
and $M$ vector spaces on field $F$. $\hat{A}\otimes{\hat{B}}$ is the kronecker product of two
operators in the space $L\otimes{M}$. Througout this article, we will use square bracets to
write the matrix form of operator and their dimensions with subscript. 
Kronecker product of two matrices are given by the rule
\begin{eqnarray}
[A]_{m1,n1}\otimes{[B]}_{m2,n2}=[C]_{(m1,m2),(n1,n2)}
\end{eqnarray} 
After the kronecker product the dimensions of the finite space becomes $N\cdot{M}$, where $N$ and
$M$ are dimensions of finite spaces on which operators $A$ and $B$ are defined. 
\section{Properties of the kronecker products}
Let the operators $\hat{A}1$, $\hat{A}2$, $\hat{B}1$, $\hat{B}2$ and $\hat{E}$ (unity matrix)
are defined in finite dimensional veector spaces $L1$, $L2$, $M1$, $M2$ and $\hat{E}$ is defined
in $L1$, $L2$, $M1$, $M2$ on field $F$. The properties of the kronecker product can be written as [9]
\begin{enumerate}
\item $[A1]_{m1,n1}\otimes{0}=0\otimes{[B1]_{m1,n1}}=0$ (where $0$ is zero matrix)
\item $[E]_{m1,n1}\otimes{[E]_{m2,n2}}=[E]_{(m1,n1),(m2,n2)}$
\item $([A1]_{m1,n1}+[A2]_{m2,n2})\otimes{[B1]_{m1,n1}}=[A1]_{m1,n1}\otimes{[B1]_{m1,n1}}+[A2]_{m2,n2}\otimes{[B1]_{m1,n1}}$
\item $[A1]_{m1,n1}\otimes{([B1]_{m1,n1}+[B2]_{m2,n2})}=[A1]_{m1,n1}\otimes{[B1]_{m1,n1}}+[A1]_{m1,n1}\otimes{[B2]_{m2,n2}}$
\item $s\cdot{[A1]_{m1,n1}}\otimes{t\cdot{[B1]_{m1,n1}}}=s\cdot{t}\cdot{[A1]_{m1,n1}\otimes{[B1]_{m1,n1}}}$ ($s, t$ are constants)
\item $([A1]_{m1,n1}\otimes{[B1]_{m1,n1}})^{-1}=[B1]_{m1,n1}^{-1}\otimes{[A1]_{m1,n1}^{-1}}$
\item $([A1]_{m1,n1}\cdot{[B1]_{n1,m1}})\otimes{([A2]_{m2,n2}\cdot{[B2]_{n2,m2}})}=([A1]_{m1,n1}\otimes{[A2]_{m2,n2}})\cdot{([B1]_{n1,m1}\otimes{[B2]_{n2,m2}})}$
\item $([A1]_{m1,n1}\otimes{[B1]_{m1,n1}})\neq([B1]_{m1,n1}\otimes{[A1]_{m1,n1}})$
\end{enumerate}
We will prove only one property (8), which is used frequently in quantum information. 
The others can be proved easily by analyzing the proof of property (8).\\
\begin{bf}Theorem:\end{bf}The kronecker product of two matrices are non-commutative i.e.,$([A]\otimes{[B]})\neq([B]\otimes{[A]})$.\\
\begin{bf}Proof:\end{bf}
Let two linear operators $\hat{A}$ and $\hat{B}$ with bases $a$ and $b$ are 
defined in finite vector spaces $L$ and $M$ on field $F$. The operators can 
be written into the form of matrices $[A]$ and $[B]$.\\
First we will write the L.H.S part of the kronecker product i.e., $[A]\otimes{[B]}$ and then R.H.S $[B]\otimes{[A]}$. 
Then, we will comare all the elements of both the L.H.S and R.H.S matrices. If even one of the element with same indicies of both the matrices
differ then these matrices are not equal or non-commutative. 
\begin{eqnarray}
([A]_{m\times{n}}\otimes{[B]_{m\times{n}}})=\left(
\begin{array}{cccc}
A_{1,1}\cdot{[B]_{m\times{n}}} & A_{1,2}\cdot{[B]_{m\times{n}}} & \cdots & A_{1,n}\cdot{[B]_{m\times{n}}}\\
A_{2,1}\cdot{[B]_{m\times{n}}} & A_{2,2}\cdot{[B]_{m\times{n}}} & \cdots & A_{2,n}\cdot{[B]_{m\times{n}}}\\
\cdots & \cdots & \cdots & \cdots\\
A_{m,1}\cdot{[B]_{m\times{n}}} & A_{m,2}\cdot{[B]_{m\times{n}}} & \cdots & A_{m,n}\cdot{[B]_{m\times{n}}}\\
\end{array}
\right)
\end{eqnarray}

\begin{eqnarray}
([B]_{m\times{n}}\otimes{[A]_{m\times{n}}})=\left(
\begin{array}{cccc}
B_{1,1}\cdot{[A]_{m\times{n}}} & B_{1,2}\cdot{[A]_{m\times{n}}} & \cdots & B_{1,n}\cdot{[A]_{m\times{n}}}\\
B_{2,1}\cdot{[A]_{m\times{n}}} & B_{2,2}\cdot{[A]_{m\times{n}}} & \cdots & B_{2,n}\cdot{[A]_{m\times{n}}}\\
\cdots & \cdots & \cdots & \cdots\\
B_{m,1}\cdot{[A]_{m\times{n}}} & B_{m,2}\cdot{[A]_{m\times{n}}} & \cdots & B_{m,n}\cdot{[A]_{m\times{n}}}\\
\end{array}
\right)
\end{eqnarray}
The elements of matrices (2) and (3) are not equal. It means the (2) and (3) are 
non-commutative.\\
\begin{bf}Note: \end{bf}Only the kronecker product of two unity matrices are equal 
i.e., they are commutative.

\section{Similar operators (matrices)}
Let two linear operators $\hat{A}$ and $\hat{B}$ with bases $a$ and $b$ are defined in vector spaces 
$L$ and $M$ on field $F$. The question arises, when operators $\hat{A}$ and $\hat{B}$ are considered similar.
Since, we are interested in the similarity of these operators, so we will study
the action of these operators on different bases in different vector spaces.\\
\begin{bf}Defination:\end{bf}
The opearators $\hat{A}:L\rightarrow{L}$ and $\hat{B}:M\rightarrow{M}$ are called similar operators,
if they are defined on field $F$, $dim{\hat{A}}=dim{\hat{B}}$ and exist isomorphism $\hat{f}:L\rightarrow{M}$,
i.e., $\hat{B}(b)=\hat{f}\hat{A}\hat{f}^{-1}(b)$.\\
\begin{bf}Theorem:\end{bf}
If linear operators $\hat{A}$ and $\hat{B}$ with bases $(a)$ and $(b)$ 
are defined in vector spaces $L$ and $M$ on field $F$ then the matrices
of operators $\hat{A}$ and $\hat{B}$ in their corresponding bases $a$ 
and $b$ are similar i.e., $[A]^{a}=[B]^{b}$.\\
\begin{bf}Proof:\end{bf}
Let the bases $a:a_{1},\ldots ,a_{n}$ and $b:\hat{f}(L\rightarrow{M})(a)=a_{1}^{'},\ldots ,a_{n}^{'}$
of linear operators $\hat{A}$ and $\hat{B}$ are defined in vector spaces $L$ 
and $M$ on field $F$ then 
\begin{eqnarray}
& \hat{A}(a_{j}) &=\sum_{i}A_{i,j}a_{i}.\\
& \hat{B}(b_{j}) &=\hat{B}\hat{f}(a_{j})=\hat{f}\hat{A}(a_{j})\nonumber\\
& &=\hat{f}\sum_{i}A_{i,j}(a_{i})=\sum_{i}A_{i,j}\hat{f}(a_{i})=\sum_{i}A_{i,j}(a_{i}^{'}).
\end{eqnarray}
Hence, $A$ and $B$ are similar matrices.\\
\begin{bf}The more simplified proof of this theorem is:\end{bf}\\
\begin{bf}Lemma:\end{bf}
Let the linear operator $\hat{C}$ defined in vector space $L$ and $M$ changes the 
bases a into b i.e.,
\begin{eqnarray}
b_{j}=\sum_{i}C_{i,j}a_{i}.\\
\hat{A}a_{j}=\sum_{i}A_{i,j}a_{i}.
\end{eqnarray}
The operartor $\hat{A}$ acts on basis $b$ gives the matrix $D_{i,j}$ and basis 
vector $b_{j}$
\begin{eqnarray}
\hat{B}b_{j}=\sum_{i}B_{i,j}b_{i}.\\
\hat{A}b_{j}=\sum_{i}D_{i,j}b_{i}.
\end{eqnarray}
By putting (3) into L.H.S of (6)
\begin{eqnarray}
\hat{A}\sum_{k}C_{k,j}a_{k}=\sum_{k}C_{k,j}\hat{A}a_{k}=\sum_{k}\sum_{i}A_{i,k}C_{k,j}a_{i}.
\end{eqnarray}
By putting (3) into R.H.S of (6)
\begin{eqnarray}
\sum_{k}D_{k,j}b_{k}=\sum_{k}D_{k,j}\sum_{i}C_{i,k}a_{i}=\sum_{k}\sum_{i}C_{i,k}D_{k,j}a_{i}.
\end{eqnarray}
The R.H.S of (7) and (8) are equal\\
\begin{eqnarray}
C_{i,k}D_{k,j}=A_{i,k}C_{k,j}.
\end{eqnarray}
or
\begin{eqnarray}
[D]=[C]^{-1}\cdot{[A]\cdot{[C]}}.
\end{eqnarray}
The theorem is proved.

\section{Physical aspects of kronecker product}
The kronecker product in group theory is widely used, especially with 
Wigner D-function [7,5,16,18]. The main purpose of its use
in physics is to get the higher dimensional vector space. For example,
in atomic physics, when we want to calculate the eigenvalues and 
eigenvectors of a system of spins $1/2$ or spin Hamiltonian. We analytically or with the help of computer diagonalize 
spin Hamiltonian and find eigenvectors and eigenvalues with two methods:\\
\begin{enumerate}
\item We should numerate each operator 
(matrix) of corresponding spin without multiplying (ordinary matrix 
multiplication) them with each other. By doing this, we can label each 
matrix of corresponding spin and each operater acts on their corresponding 
eigenvector.\\
This technique can be applied for a few number of spins. But when the 
number of spins increases this method will give only complicated calculations,
which could take a lot of time to get the result.\\ 

\item By appling the kronecker product between differnt spins matrices e.g.,
two matrices (dimensions $2\times{2}$) of spins $1/2$, we wil get the matrix 
of dimensions ($4\times{4}$). This method is very compact, which means we can 
use the computer to get the eigenvectors and eigenvalues of matrix after 
applying kronecker product for higher number of spins e.g., for the system
of spins $1/2$.\\
But there are some mathematical and physical problems during the process of 
kronecker product. The problems are 
\begin{enumerate}
\item As it is seen from the non-commutative nature of kronecker product 
that we do not have right to take kronecker product for two different spins 
freely (because they are non-commutative). Then how kronecker product can be
applied in quantum theory.\\
\item The non-commutative matrices ($[A]\otimes{[B]}\neq{[B]\otimes{[A]}}$) 
after the kronecker product are called similar matrices. It means, the 
eigenvalues of matrix $[AB]=[A]\otimes{[B]}$ and $[BA]=[B]\otimes{[A]}$ 
are similar. But eigenvectors of some eigenvalues are misplaced with their 
neighbour eigenvectors. This misplacement can be removed by applying the
smilar matrix method, which is proved earlier.\\
The similar matrix method becomes more complicated as the dimensions of the 
vector space increases (number of spins increases).
\end{enumerate}
\end{enumerate}
All of these problems, will be answered in paragraph 6.
\section{Applications in quantum theory}
At the moment the kronecker product is extensively used in quantum information [4,8,11]
theory. So, we will concentrate on the application of kronecker product in 
quantum information. All the applications of kronecker kronecker product in 
quantum information theory can be easily applied to any other branch of 
quantum theory where it requires.\\
\subsection{Examples of kronecker product}
\subsubsection{Hamiltonian of $n$ spins $1/2$ in Nuclear Magnetic Resonance}
Let $\hat{\sigma}_{1}, \hat{\sigma}_{2}, \cdots, \hat{\sigma}_{n}$ are linear spin operators 
defined in the finite dimensional $S_{1}, S_{2}, \cdots, S_{n}$ vector spaces on field $F$. 
$\hat{\sigma}_{1}\otimes{\hat{\sigma}_{2}}\cdots{\otimes{\hat{\sigma}_{n}}}$ are defined in linear space $S_{1}\otimes{S_{2}}\cdots{\otimes{S_{n}}}$. 
All the matrices of spin operators $\hat{\sigma}_{1}, \hat{\sigma}_{2}, \cdots, \hat{\sigma}_{n}$ are $2\times{2}$ dimensions.
For simplicity, we are taking $\hbar=1$.
\subsubsection{When $n=2$ spins $1/2$}
Hamiltonian of two spins $\hat{\sigma}_{1}$ and $\hat{\sigma}_{2}$ defined in
linear space $S_{1}, S_{2}$. $\hat{\sigma}_{1z}$, $\hat{\sigma}_{2z}$ coupling 
with hyperfine interaction $J_{12}$ are placed parallel to applied constant 
magnetic field $B_0\|$$z-axis$:
$$
\hat{H}2=-\mu{B_0}\cdot{(\hat{\sigma}_{z1}\otimes{\hat{E}_2}})-\mu{B_0}\cdot{(\hat{E}_{1}\otimes{\hat{\sigma}_{z2}}})
+J_{12}(\hat{\sigma}_{x1}\otimes{\hat{\sigma}_{x2}})
+J_{12}(\hat{\sigma}_{y1}\otimes{\hat{\sigma}_{y2}})
+J_{12}(\hat{\sigma}_{z1}\otimes{\hat{\sigma}_{z2}})
$$
\subsubsection{When $n=3$ spins $1/2$}
Hamiltonian of three spins $\hat{\sigma}_{1}$, $\hat{\sigma}_2$ and $\hat{\sigma}_{3}$ 
are defined in linear space $S_{1}, S_{2}, S_{3}$ and coupling with hyperfine interaction 
$J_{12}$ between $\hat{\sigma}_{1}$ and  $\hat{\sigma}_{2}$, $J_{23}$ between $\hat{\sigma}_{2}$
and $\hat{\sigma}_{3}$ and $J_{31}$ between spins $\hat{\sigma}_{3}$ and
$\hat{\sigma}_{1}$. $\hat{\sigma}_{1z}, \hat{\sigma}_{2z}$ and $\hat{\sigma}_{3z}$ are placed 
parallel to applied constant magnetic field $B_0\|$$z-axis$:

\begin{eqnarray*}
\hat{H}3=-\mu{B_0}\cdot{(\hat{\sigma}_{z1}\otimes{\hat{E}_2}}\otimes{\hat{E}_3})
-\mu{B_0}\cdot{(\hat{E}_{1}\otimes{\hat{\sigma}_{z2}}}\otimes{\hat{E}_{3}})\nonumber \\
-\mu{B_0}\cdot{(\hat{E}_{1}\otimes{\hat{E}_{2}}}\otimes{\hat{\sigma}_{z3}})
+J_{12}(\hat{\sigma}_{1}\otimes{\hat{\sigma}_{2}}\otimes{\hat{E}_{3}})\nonumber \\
+J_{23}(\hat{E}_{1}\otimes{\hat{\sigma}_{2}}\otimes{\hat{\sigma}_{3}})
+J_{31}(\hat{\sigma}_{3}\otimes{\hat{\sigma}_{1}}\otimes{\hat{E}_{2}})
\end{eqnarray*}
Hamiltonians of higher number of spins $1/2$ can be written in the same way as for $2$
and $3$ spins $1/2$.\\
\subsection{Kronecker product in quantum information theory to get
the spin Hamiltonians}
To write the spin Hamiltonian, first of 
all we should write $\hat{S}_{x}$, $\hat{S}_{y}$, $\hat{S}_{z}$, $\hat{S}^{2}$ and 
then add them with their coresponding factors, we will get the spin Hamiltonians (e.g. $\hat{H}2$ and $\hat{H}3$).
We will proof this later.\\
Let we want to write the spin Hamiltonian of $n$ nuclear spins in NMR (Nuclear Magnetic Resonance):\\ 
\begin{enumerate}
\item Total projection of $n$ spins $1/2$ on z-axis is conserved.
\begin{eqnarray}
\hat{S}_{z}=1/2(\hat{\sigma}_{1z}
\left\{
\begin{array}{cc}
\otimes^{n}_{i=2}\hat{E}_{i}, & \mbox{if } n\ge{2}\\
1 & \mbox{if } n=1.
\end{array}
\right\}
+\hat{E}_{1}\otimes{\hat\sigma}_{2z}
\left\{
\begin{array}{cc}
\otimes^{n}_{i=3}\hat{E}_{i}, & \mbox{if } n\ge{3}\\
1 & \mbox{if } n=2\\
0 & \mbox{if } n<2.
\end{array}
\right\}\nonumber \\
+\hat{E}_{1}\otimes{\hat{E}}_{2}\otimes{\hat{\sigma}_{3z}}
\left\{
\begin{array}{cc}
\otimes^{n}_{i=4}\hat{E}_{i}, & \mbox{if } n\ge{3}\\
1 & \mbox{if } n=3\\
0 & \mbox{if } n<3.
\end{array}
\right\}
+\cdots )
\end{eqnarray}
The first term in (14) contains first spin $\hat{\sigma}_{1z}$ 
with kronecker product of $\hat{E}_{i}$ unit matrices of other spins i.e., second,
third and so on. The second term contains first factor unit matrix $\hat{E}_{1}$ with
kronecker product of second spin $\hat{\sigma}_{2z}$ and unit matrix of other spins. 
The third and forthcoming terms are written analytically by analyzing precedings.\\

$\hat{S}_{x}$ and $\hat{S}_{y}$ can be written by putting 
$\hat{\sigma}_{x}$ and $\hat{\sigma}_{y}$ in place of $\hat{\sigma}_{z}$.
\item The square of the total spin
$
\hat{S}^2=\hat{S}\cdot{(\hat{S}+1)}
$ is conserved.
\begin{equation}
\hat{S}=i\hat{S}_{x}+j\hat{S}_{y}+k\hat{S}_{z}\nonumber \\
\end{equation}
\begin{equation}
\label{math/2}
\hat{S}^2=\hat{S}_{x}^{2}+\hat{S}_{y}^{2}+\hat{S}_{z}^{2}
\end{equation}
\item Equations (14) and (16) are constant of motion.
That is $[\hat{S}_{z},\hat{S}^{2}]=0$. It means that the eigenvalues and 
eigenvectors of (14) and (16) are identicals.
\item The Hamiltonians $\hat{H}2$ and $\hat{H}3$ are consist of two parts
(14) and (16) with their corresponding factors.
\item \begin{bf}Proof of $\hat{H}2$\end{bf}\\
To proof $\hat{H}2$, that it consists of (14) and (16), we should take two 
spins $\hat{S}_{1}$ and $\hat{S}_{2}$ with constants $a_{i}$, $i\in{x,y,z}$.
To write $S_{x}$, $S_{y}$ and $S_{z}$, we use (14) i.e.,
\begin{equation}
\hat{S}_{z}=
a_{z}(\hat{\sigma}_{1z}\otimes{\hat{E}_{2}}+\hat{E}_{1}\otimes{\hat{\sigma}_{2z})}
\end{equation}
\begin{equation}
\hat{S}^{2}=\hat{S}_{x}^{2}+\hat{S}_{y}^{2}+\hat{S}_{z}^{2}
\end{equation}
where
\begin{eqnarray}
\hat{S}_{x}^{2}&= & (\hat{\sigma}_{1z}\otimes{\hat{E}_{2}})(\hat{E}_{1}\otimes{\hat{\sigma}_{2z}})+(\hat{\sigma}_{1z}\otimes{\hat{E}_{2}})(\hat{E}_{1}\otimes{\hat{\sigma}_{2z}})\nonumber \\
& &+(\hat{\sigma}_{1x}\otimes{\hat{E}_{2}})(\hat{\sigma}_{1x}\otimes{\hat{E}_{2}})+(\hat{\sigma}_{1x}\otimes{\hat{E}_{2}})(\hat{E}_{1}\otimes{\hat{\sigma}_{2x}})\nonumber \\
& &+(\hat{E}_{1}\otimes{\hat{\sigma}_{2x}})(\hat{\sigma}_{1x}\otimes{\hat{E}_{2}})+(\hat{E}_{1}\otimes{\hat{\sigma}_{2x}})(\hat{E}_{1}\otimes{\hat{\sigma}_{2x}})
\end{eqnarray}
By using the property (7), we can simplified (19) to
\begin{eqnarray}
\hat{S}_{x}^{2} & = & 2a_{x}^{2}[(\hat{E}_{1}\otimes{\hat{E}_{2}})+(\hat{\sigma}_{1x}\otimes{\hat{\sigma}_{2x}})]
\end{eqnarray}
Similarly, we can calculate $\hat{S}_{y}^{2}$ and $\hat{S}_{z}^{2}$
\begin{eqnarray}
\hat{S}_{y}^{2} & = & 2a_{y}^{2}[(\hat{E}_{1}\otimes{\hat{E}_{2}})+(\hat{\sigma}_{1y}\otimes{\hat{\sigma}_{2y}})]\\
\hat{S}_{z}^{2} & = & 2a_{z}^{2}[(\hat{E}_{1}\otimes{\hat{E}_{2}})+(\hat{\sigma}_{1z}\otimes{\hat{\sigma}_{2z}})]
\end{eqnarray}
Now we will add (17), (20), (21) and (22)
\begin{eqnarray}
a_{z}(\hat{\sigma}_{1z}\otimes{\hat{E}_{2}}+\hat{E}_{1}\otimes{\hat{\sigma}_{2z})}\nonumber\\
+2a_{x}^{2}[(\hat{E}_{1}\otimes{\hat{E}_{2}})+(\hat{\sigma}_{1x}\otimes{\hat{\sigma}_{2x}})]\nonumber\\
+2a_{y}^{2}[(\hat{E}_{1}\otimes{\hat{E}_{2}})+(\hat{\sigma}_{1y}\otimes{\hat{\sigma}_{2y}})]\nonumber\\
+2a_{z}^{2}[(\hat{E}_{1}\otimes{\hat{E}_{2}})+(\hat{\sigma}_{1z}\otimes{\hat{\sigma}_{2z}})]
\end{eqnarray}
By analyzing (23) and $\hat{H}2$, we can see that they are same, only we have to chosse
corresponding $a_{i}$. The term $2a_{i}^{2}\hat{E}_{1}\otimes{\hat{E}_{2}}$ does not have any meaning, 
since it is unit matrix. We have proved that $\hat{H}2$ can be correctly chosen with the help of
$\hat{S}_{z}$ and $\hat{S}^{2}$.\\
\begin{bf}Note:\end{bf}
\begin{eqnarray}
\hat{\sigma}_{1x}\otimes{\sigma}_{2x}=\hat{\sigma}_{2x}\otimes{\sigma}_{1x}\nonumber\\
\hat{\sigma}_{1y}\otimes{\sigma}_{2y}=\hat{\sigma}_{2y}\otimes{\sigma}_{1y}\nonumber\\
\hat{\sigma}_{1z}\otimes{\sigma}_{2z}=\hat{\sigma}_{2z}\otimes{\sigma}_{1z}
\end{eqnarray}
The proof of $\hat{H}3$ can be written similarly by applying property (7). 
\end{enumerate}
\section{Conclusion}
We have proposed the method by which one can get the exact spin Hamiltonian. Also, 
it is shown that we do not have right to take kronecker product freely i.e., left/right to 
any operator, since the kronecker product is non-commutative. The proposed method has applied to
get the spin Hamiltonian to the case of NMR. The proposed method can be used to get the higher 
number of spins, which is very important in quantum information.

\end{document}